\documentclass[twocolumn,showpacs]{revtex4}

\usepackage{graphicx}
\usepackage{dcolumn}
\usepackage{bm}

\newcommand{\ket}[1]{\left|#1\right\rangle}
\newcommand{\bra}[1]{\left\langle#1\right|}

\newcommand{\be}{\begin{equation}}
\newcommand{\ee}{\end{equation}}

\newcommand{\dx}{{\rm d}{\bf x}}
\newcommand{\dr}{{\rm d}{\bf r}}
\newcommand{\bx}{{\bf x}}
\newcommand{\br}{{\bf r}}
\newcommand{\bq}{{\bf q}}

\newcommand{\ke}[1]{|#1\rangle}

\newcommand{\da}{^\dagger}
\newcommand{\pt}[1]{\left( #1 \right)}
\newcommand{\pq}[1]{\left[ #1 \right]}
\newcommand{\pg}[1]{\left\{ #1 \right\}}

\newcommand{\al}[1]{^{(#1)}}

\newcommand{\lpg}[1]{\left\{ #1 \right.}

\newcommand{\rpg}[1]{\left. #1 \right\}}

\begin{document}

\title{Quantum-noise quenching in atomic tweezers}%
\author{Stefano Zippilli,$^{1,2,3}$ Bernd Mohring,$^4$ Eric Lutz,$^5$ Giovanna Morigi,$^{1,2}$ and Wolfgang Schleich$^4$}
\affiliation{$^1$ Departament de F\'{\i}sica, Universitat Aut\`onoma de Barcelona,
E-08193 Bellaterra, Spain\\
$^2$ Theoretische Physik, Universit\"at des Saarlandes, D-66041 Saarbr\"ucken, Germany\\
$^3$ Fachbereich Physik and research center OPTIMAS, Technische Universit\"at Kaiserslautern, D-67663 Kaiserslautern, Germany\\
$^4$ Institut f\"ur Quantenphysik, Universit\"at Ulm, D-89081 Ulm, Germany\\
$^5$ Department of Physics, University of Augsburg, D-86135 Augsburg, Germany
}
\date{\today}

\begin{abstract}
The efficiency of extracting single atoms or molecules from an ultracold bosonic reservoir is theoretically investigated for a protocol based on lasers, coupling the hyperfine state in which the atoms form a condensate to another stable state, in which the atom experiences a tight potential in the regime of collisional blockade, the quantum tweezers. The transfer efficiency into the single-atom ground state of the tight trap is fundamentally limited by the collective modes of the condensate, which are thermally and dynamically excited.  The noise due to these excitations can be quenched for sufficiently long laser pulses, thereby achieving high efficiencies. These results show that this protocol can be applied for initializing a quantum register based on tweezer traps for neutral atoms. \end{abstract}


\maketitle

Optical tweezers hold and manipulate microparticles, from molecules to living cells~\cite{Ashkin,Ritort,Rubinzstein-Dunlop}. The concept on which they are based find applications down to the level of single atoms~\cite{Meschede,Diener,Grangier}. Indeed, the progress in the mechanical manipulation of atoms by means of lasers has allowed, amongst others, to control the position and transport of cold particles loaded in dispersive potentials~\cite{Meschede,Dumke,Grangier,Bohi}. Such control is at the basis of several protocols for quantum information processing with neutral atoms~\cite{Calarco00,Calarco,Hayes,Rydberg}. In this context a relevant issue is the initialization of the quantum register~\cite{Gruenzweig,Gehr,Serwame}, namely, their preparation in target quantum states of the single-atom trap, the quantum tweezers.

\begin{figure}[h!]
  \begin{center}
   \includegraphics[width=8cm]{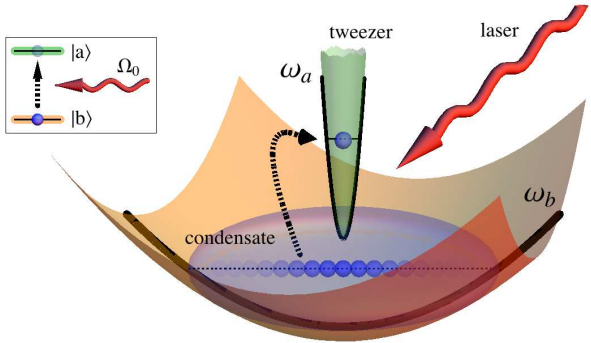}
  \end{center}
\caption{(Color online) %
An atom is transferred from a Bose-Einstein condensate (prepared in the electronic state $\ket{b}$) to the ground state of a tweezer trap (realized when the atom is in state $\ket a$) by a Raman or microwave transition coherently coupling the two stable states. The protocol is based on spectrally resolving the one-atom ground state of the tweezer trap, which is in the collisional blockade regime~\cite{Mohring}. The efficiency of the extraction protocol can be enhanced by quenching the quantum noise due to the condensate excitations. }%
\label{Fig:1} \end{figure}

Proposals for extracting atoms on demand from a quantum reservoir are based either on tunneling, and/or dynamical modification of the trapping potential~\cite{Diener,Culling,Raizen,Kollath}. Another approach utilizes electromagnetic radiation for transferring a single atom from a condensate into the ground state of a tweezer trap, and achieves it by coupling two internal atomic states with different spin-dependent potentials~\cite{Rabl,Mohring,Jaksch,Recati05}. The basic idea is that the ground state of the tweezers trap can be spectrally resolved in the collisional blockade regime, so that the single-atom ground state can be coupled on resonance with the ultracold reservoir, while all other states of the quantum tweezers are set significantly out of resonance. This approach usually discards the effect of the condensate excitations~\cite{Mohring,Rabl}, which may instead be significant in various setups where the condensate is realized in relatively shallow trapping potentials. These excitations are present at finite temperature and non-vanishing interactions. They are also created by the dynamics of the extraction process and the collisions between the atom in the tweezer trap and the condensate. They are hence a source of quantum noise, which is inherently due to matter-wave fluctuations and which is expected to reduce the efficiency of the protocol.

In this article we show that quantum noise due to matter-wave fluctuations can be quenched in atomic tweezers for accessible experimental parameters. Quenching of noise is achieved by means of a destructive interference between dynamics of different physical origin. Our findings are in agreement with and generalize the predictions in Ref.~\cite{Recati05}, which have been derived under specific assumptions. These results show that this procedure can be used for initializing quantum registers based on tweezers arrays~\cite{Grangier,Raizen}.

The setup we consider is sketched in Fig.~\ref{Fig:1}. Here, a coherent Raman (or microwave) transition couples two internal, stable states of the atoms, $\ket b$ and $\ket a$, in which the atoms experience a shallow and a steep confining potential, respectively. Transitions between the two states allow one for switching between the two confinements. The atoms are identical bosons of mass $M$ and form a Bose-Einstein condensate in state $\ket b$, i.e., the quantum reservoir. In absence of perturbations their dynamics is described by Hamiltonian ${\cal  H}_b=\int \dr\,  \psi_{b}\da(\br)\pq{-\frac{\hbar^2}{2M} \nabla^2+V_b(\br)+(g_b/2) \psi_b\da(\br)\psi_{b}(\br) } \psi_{b}(\br)$, with $V_b({\bf r})$ the potential, $\psi_b$ and $\psi_b^{\dagger}$ the bosonic field operators annihilating and creating an atom in state $\ket{b}$ at position $\br$, and $g_b=4\pi\hbar^2a_{b}/M$ the interaction strength of two-body, $s$-wave collisions with scattering length $a_{b}$. A radiation pulse couples state $\ket b$ to $\ket a$, in which the atomic center of mass is confined by the steep potential $V_a(\br)$ in the regime of collisional blockade, i.e., the quantum tweezers. We denote by $\psi_a(\br)$ and $\phi_a(\br)$ the field operator and wave function of the atom in the ground state of the tweezer trap, here assumed to be harmonic. The pulse is a standing wave with wave vector ${\bf k}$ and is homogeneous over the volume of the tweezers. It has duration $\tau$, characteristic frequency $\omega_L$ and maximum value of the Rabi frequency $\Omega_0$. The Hamiltonian for the dynamics of atom-light coupling reads ${\cal H}_{\rm int}(t)={\cal H}_{\rm r}(t)+{\cal H}_{\rm off}(t)+{\cal H}_{\rm c}$, where
\begin{equation}
\label{h:res}
{\cal H}_{\rm r}(t)=\frac{\hbar\Omega_0}{2}f(t)\int \dr\, \cos({\bf k}\cdot \br) \psi_{a}\da(\br)\psi_{b}(\br){\rm e}^{-{\rm i}\omega_Lt}+{\rm H.c.}
\end{equation}
describes the resonant coupling between the condensate and the single-atom ground state in the tweezers, with $f(t)$ the temporal shape of the pulse, here assumed to be a step function.  ${\cal H}_{\rm off}(t)$ includes the coupling to all other bound states of the tweezers, and ${\cal H}_{\rm c}=(g_{ab}/2)\int \dr\, \psi_{b}\da(\br)\psi_a\da(\br)\psi_{a}(\br) \psi_{b}(\br)$ describes $s$-wave collisions with strength $g_{ab}$ between atoms in $\ket b$ and $\ket a$. In the regime of collisional blockade the frequency $\omega_{\rm gap}\sim (g_{a}/2\hbar)\int {\rm d}\br \, |\phi_a(\br)|^4 $ gives the gap between the single- and the two-atom ground state in the tweezers, with $g_a$ the strength of interparticle collisions in $\ket a$. For a harmonic potential $V_a(\br)$ with frequencies of the order of hundreds of kHz till MHz, $\omega_{\rm gap}$ can reach the order of several to hundreds kHz, and the gap can be spectrally resolved~\cite{Diener,Rabl,Mohring,Recati05}. In this regime the laser resonantly couples the condensate with the single-atom ground state of the tweezers with strength $\Omega_{\rm eff}=\Omega_0\int \dr\, \cos({\bf k}\cdot\br)\phi_a(\br)\phi_b(\br)$, while the dynamics due to ${\cal H}_{\rm off}$ can be neglected. This requires $\omega_{\rm gap}\tau\gg 1$ and $\Omega_{\rm eff}\ll \omega_{\rm gap}$. Correspondingly, the Hilbert space of the tweezers is reduced to the states ${\ket 0}_a$ and ${\ket 1}_a$, i.e., no atoms and one atom in the tweezers ground state, respectively. It is convenient to define the operator $\sigma=|0\rangle_a\langle 1|$, such that $\psi_a(\br)= \phi_a(\br)\sigma$.

We now focus on the effect of the condensate excitations over the efficiency of the extraction dynamics. In the following we assume that the probability of populating non-condensed states during the extraction process is small. For sufficiently low temperatures the field operator for the atoms in the condensate can be decomposed into the sum~\cite{Stringari_Book}
\begin{equation}\label{Bog}\psi_b(\br)= \phi_b(\br)+\delta\psi_b(\br)\,,\end{equation} where $\phi_b(\br)$ is the macroscopic wave function of the condensate, satisfying the Gross-Pitaevskii equation $\left(-\frac{\hbar^2\nabla^2}{2M}+V_b(\br)+g_b|\phi_b|^2\right)\phi_b(\br)=\mu\phi_b(\br)$ with $\mu$ the chemical potential. Operator $\delta\psi_b(\br)$ represents the quantum fluctuations about the mean value, and in the Bogoliubov expansion reads $\delta\psi_b(\br)=\sum_\bq\left[u_\bq(\br)b_\bq-v^\ast_\bq(\br)b_\bq^\dagger\right]\,,$ with $b_\bq$ and $b_\bq\da$ the annihilation and creation operators, respectively, of a quasiparticle with frequency $\omega_{\bq}$, and $u_\bq(\br)$ and $v_\bq(\br)$ the corresponding wave functions, such that the Hamiltonian for the atoms in state $\ket b$ reads ${\cal H}_{b}\simeq\hbar \sum_\bq \omega_\bq\, b_\bq^\dagger b_\bq\,$. The total system dynamics is thus mapped to a spin-boson model~\cite{Leggett,Recati05}, where the bosonic bath are the Bogoliubov excitations of the condensate and the spin is composed by the tweezers states ${\ket 0}_a$ and ${\ket 1}_a$, eigenstates of the Pauli matrix $\sigma_z$. Using Eq.~(\ref{Bog}) into Eq.~(\ref{h:res}), the coupling between condensate and ground state of the tweezers is given by Hamiltonian $\mathcal{H}_{s}= \hbar\Omega_{\rm eff}f(t)\sigma_x/2$, while the coupling involving the Bogoliubov excitations, which emerges from the corresponding terms in ${\cal H}_{\rm r}+{\cal H}_{\rm c}$, reads
\begin{eqnarray}\label{HSR}
\mathcal{H}_{sb}=\frac{\hbar}{2}\sum_{\bf q}\left(\alpha_{x,{\bf q}}(t)\sigma_x+{\rm i}\alpha_{y{\bf q}}(t)\sigma_y+2\alpha_{z{\bf q}}\sigma_z\right)b_{\bf q}+{\rm H.c.}
\end{eqnarray}
Here, the first two terms on the right hand side originate from the coupling of the pseudospin to the bosonic bath via the laser, with coupling strengths
\begin{eqnarray}\label{alpha:1}
 \alpha_{x{\bf
q}}(t)&=&\frac{\Omega_0}{2}f(t)
\int \dx\, \cos({\bf k}\cdot \bx) \phi_a(\bx) \pq{u_\bq(\bx)-v_\bq(\bx)}\nonumber \\
\alpha_{y{\bf q}}(t)&=&\frac{\Omega_0}{2}f(t)\int \dx\, \cos({\bf k}\cdot \bx) \phi_a(\bx) \pq{u_\bq(\bx)+v_\bq(\bx)}\nonumber\\
\label{alpha:2}
\end{eqnarray}
while the third term is due to collisions between the condensate trap and the tweezers with strength
\begin{eqnarray}
\label{alpha:3}
\alpha_{z{\bf q}}=\frac{g_{ab}}{2\hbar}\int \dx\, \left|\phi_a(\bx)\right|^2 \phi_b(\bx)
\pq{u_\bq(\bx)-v_\bq(\bx)}.
\end{eqnarray}
These two kinds of perturbation couple the effective spin with the condensate excitations and may interfere~\cite{Recati05}.
\begin{figure*}[ht!]
  \begin{center}
      \includegraphics[width=18cm]{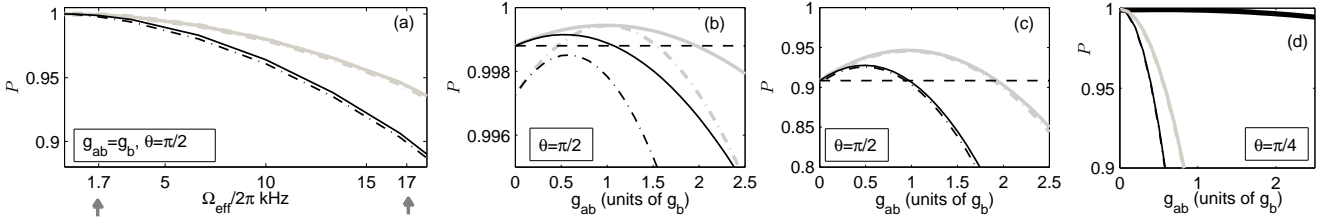}
\caption{(a) Efficiency $P$ of preparing the tweezers in state $\ket{1}_a$ ($\theta=\pi/2$) as a function of $\Omega_{\rm eff}$ and at the corresponding time $\tau=\tau_0$, for  $g_{ab}=g_b$. The arrows indicate the values of $\Omega_{\rm eff}$ used in plots (b) and (c). In plot (b) $P$ is displayed at $\Omega_{\rm eff}=2\pi\times1.7$ kHz and in~(c) at $\Omega_{\rm eff}=2\pi\times17$ kHz as a function of the interparticle collision strength $g_{ab}$ (in units of $g_b$). (d): same as (b) but for state $\theta=\pi/4$. The reservoir is here a condensate of $^{87}$Rb atoms in a spherical harmonic trap with $\omega_b=2\pi\times 200$ Hz and density at the trap center $n=2\times 10^{21} m^{-3}$  ($3\times10^6$ atoms). The tweezer trap frequency is $\omega_a=2\pi\times 1$ MHz, such that $\omega_{\rm gap}\sim 2\pi\times0.2$~MHz. The curves are evaluated at $T=0$ (solid line), $T=300$ nK (dash-dotted line) for the frequency of the Bogoliubov modes $\omega_{j,\ell}=\sqrt{2 j^2+2 j \ell+3j+\ell}$~\cite{Stringari_Book,Ohberg}: the gray (black) lines are a sum over $j\in \pq{1,500}$ and $\ell=0$ ($\ell=0,2$). The dashed horizontal line marks the efficiency when $g_{ab}=0$ and $T=0$. The solid-thick-black line in (d) corresponds to the efficiency of preparing state $\ket{1}_a$ at $T=0$ (same as solid-black line in (b)).
}\label{Fig:2}
  \end{center}
 \end{figure*}

We evaluate now the fidelity of preparing the tweezers in a target state, which we denote by ${\ket \theta}_a=\cos\theta {\ket 0}_a -{\rm i}\sin \theta {\ket 1}_a$, assuming that initially all atoms are in state $\ket b$ at temperature $T$. The density matrix at $t=0$ reads $\rho(0)=\ke{0}_a\bra{0}\otimes \rho_B$, with  $\rho_B={\rm e}^{-H_{b}/k_BT}/Z$ and $Z={\rm Tr}\pg{ {\rm e}^{- H_{b}/k_BT}}$. The fidelity can be cast in the form
\begin{eqnarray}
\label{Efficiency} P(\theta,\tau)&=&{\rm
Tr}\{|\theta\rangle\langle \theta|{\rm e}^{-{\rm i}{\cal H}_{\rm eff}\tau/\hbar}\rho(0){\rm e}^{{\rm i}{\cal H}_{\rm eff}\tau/\hbar}\}\nonumber\\&=&\cos^2\left(\theta-\Omega_{\rm eff}\tau/2\right)-g(\theta,\tau) \end{eqnarray} where ${\cal H}_{\rm eff}={\cal H}_{s}+{\cal H}_{b}+{\cal H}_{sb}$ and $g(\theta,\tau)$ is due to the Bogoliubov modes. In second order in the coupling $\mathcal H_{sb}$, maximum transfer efficiency is achieved setting $\tau=\tau_0\equiv 2\theta/\Omega_{\rm eff}$, namely, to the value at which perfect transfer is observed in the ideal dynamics. With this choice, the transfer efficiency reads $P(\theta,\tau_0)=1-g(\theta,\tau_0)$, where
\begin{eqnarray}
\label{fun:g}
g(\theta,\tau_0)&=&\pi^2\sum_{\bq}\lpg{A_{1\bq}\cos \theta
}\\&&
\rpg{+(2n_{\bq}+1)\left[A_{2\bq}\cos 2\theta+A_{3\bq}+A_{4\bq}\right]}>0\,,\nonumber
\end{eqnarray}
with $\bar n_\bq={\rm Tr}\pg{b\da_\bq b_\bq \rho_B}$ the mean thermal phonon number of the mode $\omega_{\bq}$. The other coefficients take the form
\begin{eqnarray*}
A_{1\bq}&=&-\alpha_{x\bq}\delta\al{\tau_0}\pt{\omega_\bq}\left[\alpha_{-\bq}
\delta\al{\tau_0}\pt{\omega_\bq^{-}}+\alpha_{+\bq}\delta\al{\tau_0}\pt{\omega_\bq^{+}}\right]\,,\nonumber\\
A_{2\bq}&=&\left(\frac{\alpha_{+\bq}\alpha_{-\bq}}{2}\right)\delta\al{\tau_0}\pt{\omega_\bq^{-}}\delta\al{\tau_0}\pt{\omega_\bq^{+}}\,,
\nonumber\\
A_{3\bq}&=&\frac{1}{4}\pq{
\left(\alpha_{-\bq}\delta\al{\tau_0}\pt{\omega_\bq^{-}}\right)^2+\left(\alpha_{+\bq}\delta\al{\tau_0}\pt{\omega_\bq^{+}}\right)^2}\,,
\nonumber\\
A_{4\bq}&=&\alpha_{x\bq}^2{\delta\al{\tau_0}\pt{\omega_\bq}}^2
\,,
\end{eqnarray*}
with $\delta^{(\tau)}\pt{x}={\sin\pt{x\tau/2}}/\pt{\pi x}$, $\alpha_{\pm\bq}=\alpha_{y\bq}\pm 2\alpha_{z\bq}$, and $\omega_\bq^\pm=\omega_\bq\pm\Omega_{\rm eff}$. After rewriting Eq.~(\ref{fun:g}) as a function of these coefficients, it is visible that the noise due to the laser can interfere with the one due to interspecies collisions. This occurs at sufficiently long times, for which the frequencies $\omega_\bq^\pm$ are spectrally resolved.

We now focus on the parameter regime in which quantum noise can be quenched, and study the dependence of the interference condition on the temperature $T$ and on the ``Bloch'' angle $\theta$ of the target state. At zero temperature ($n_\bq=0$) the condition on the parameters, for which function~(\ref{fun:g}) is minimal, depends on $\theta$. For the target state ${\ket \theta}_a=\pm {\rm i}{\ket 1}_a$ maximal quenching is found when
\begin{equation}
\label{condition}
 \omega_\bq\alpha_{y\bq}-2\Omega_{\rm eff}\alpha_{z\bq}\sim 0
\end{equation}
is fulfilled, for which $g_{\rm min}(\pi/2,\tau_0)=\pi^2\sum_{\bq}A_{4\bq}$. Using Eqs.~(\ref{alpha:1})-(\ref{alpha:3}) in Eq.~(\ref{condition}), one finds that the interference condition depends on the trap, density, and interparticle interaction strength, but not on $\Omega_0$. Condition~(\ref{condition}) can be further simplified in the limit in which only long-wavelength modes of the condensate are involved, and reduces to the expression $g_{ab}= g_b$, which agrees with the result derived for a low energy model in Ref.~\cite{Recati05}. For other target states one finds different conditions on the parameters, and also lower efficiencies: It results, in fact, that for $\theta\neq m\pi/2$ it is not possible to disentangle the condensate excitations from the tweezers. In particular, the state which is most sensitive to quantum noise is the one at $\theta=(2m+1)\pi/4$, with $m=0,1,2,3$: This state corresponds to the most non-classical state, equal superposition of one and zero atom in the tweezers.

The dependence of the fidelity on the choice of the target state $\theta$ is even more enhanced at finite temperatures: In this case the state which is most efficiently prepared is ${\ket \theta}_a=\pm {\rm i}{\ket 1}_a$. For this state, condition~(\ref{condition}) leads to noise quenching at finite $T$ with $g_{\rm min}(\pi/2,\tau_0)=\pi^2\sum_{\bq}(2n_{\bq}+1)A_{4\bq}$. We remark that one general physical consequence is that maximal coherence can be achieved provided that the noise due to collision between condensate and tweezers is significantly different from zero: This noise source can be tuned by means of a Feshbach resonance so to interfere destructively with the laser excitation. We also note that the function $g_{\rm min}(\pi/2,\tau_0)$ scales with the coupling strength $\Omega_0^2$, as it originates from the excitations due to the laser coupling which do not interfere with interspecies collisions. This implies that higher efficiencies are attained for lower values of $\Omega_0$, and hence for longer transfer pulses.

We now provide some examples for a condensate of $^{87}$Rb atoms in the Thomas-Fermi regime. When evaluating the fidelities we consider only long-wavelength excitations of the condensate, for which the explicit dispersion relations are known for several trap geometries~\cite{Stringari_Book}. Modes at higher energy are out of resonance and their effect can be neglected. Figure~\ref{Fig:2}(a) displays the efficiency of preparing state $\ket{1}_a$ by coupling the tweezers with a condensate in a spherical harmonic trap, when the condition $g_{ab}=g_b$ is met. The fidelity is higher when the laser pulse is sufficiently long, so to minimize the effect of the excitations created by the laser coupling which is out of phase with the interspecies collisions, and decreases with the temperature and when excitations at higher angular momentum (black lines) are included.  This latter coupling, however, can be minimized by an accurate design of the setup. Figures~\ref{Fig:2}(b) and~(c) show the dependence of the fidelity on the value of $g_{ab}$ for two specific pulse lengths: One clearly observes a maximum at $g_b\sim g_{ab}$, which lies well above the efficiency one would obtain when $g_{ab}=0$ (i.e., in absence of interspecies collisions). Figure~\ref{Fig:2}(d) shows the transfer efficiency as a function of $g_{ab}$ when the target state is $\theta=\pi/4$ and the
other parameters are as in Fig.~\ref{Fig:2}(b): the maximum is here found when $g_{ab}=0$, and lies below the efficiency found for $\theta=\pi/2$.

\begin{figure}[t!]
  \begin{center}
      \includegraphics[width=8cm]{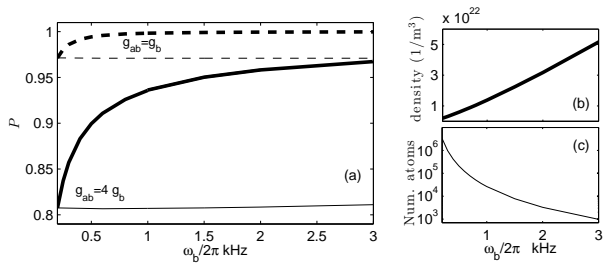}
\caption{(a) Efficiency of preparing the tweezers in state $\ket{1}_a$ as a function of $\omega_b$ for a condensate of $^{87}$Rb atoms at $T=0$ and for $g_{ab}=4 g_b$ (solid lines) and $g_{ab}=g_b$ (dashed lines). The curves are found summing up to 500 Bogoliubov modes, for either constant atomic number of atoms $~3\times10^6$ (thick lines) or constant density $n=2\times10^{21}$m$^{-3}$ (thin lines). Plots (b) and (c) display the corresponding dependence of density and atom number on $\omega_b$. The other parameters are as in Fig.~\ref{Fig:2}.}
\label{Fig:3new}
\end{center}
 \end{figure}

Figure~\ref{Fig:3new} display the efficiency of preparing the tweezers in state $\ket{1}_a$ as the trap frequency of the condensate is increased, which corresponds to approach the limit in which the condensate excitations can be spectrally resolved. The curves report two cases, where the frequency of the condensate is increased keeping constant either the atomic number (thick lines) or the density (thin lines). While in the first case the efficiency becomes significantly larger, in the second case it grows very slowly in the considered interval. We also notice that best performances are found when the condition of quenching is fulfilled at any value of $\omega_b$.

To conclude, the condensate excitations limit the efficiency of preparing quantum tweezers by loading atoms from a condensate, nevertheless their effect can be quenched by means of an interference process emerging from the dynamics induced by laser and particle-particle collisions. This requires sufficiently long transfer pulses and tuning of the various parameters, so to maximize the interference and achieving high fidelities. These concepts can be extended to protocols for creating entangled atoms in two distant tweezers traps coupled to the same condensate, developing on proposals~\cite{Klein,Recati_2}. In a more general framework, the dynamics here reported are another example of quantum reservoir engineering~\cite{QRE}, where noise may constitute a useful resource for endorsing the quantum dynamics of a physical system.

We acknowledge discussions with J. Anglin, C. Foot, M. K\"ohl, M. Raizen, J. Reichel, and P. Treutlein. Support by the European Commission (EMALI MRTN-CT-2006-035369; AQUTE), by the Spanish Ministerio de Ciencia y Innovaci\'on (QOIT, Consolider-Ingenio 2010; QNLP, FIS2007-66944), by the German Research Foundation (DFG, MO1845/1-1, LU1382/1-1) and the cluster of excellence Nanosystems Initiative Munich (NIM).

\end{document}